\documentclass[twocolumn,preprintnumbers,amsmath,amssymb,aps,prb,longbibliography]{revtex4-1}

\usepackage{graphicx}
\usepackage{calrsfs}
\usepackage{xcolor}

\begin{document}

\title{Skyrmion Molecular Crystals and Superlattices on Triangular Substrates}
\author{J. C. Bellizotti Souza$^{1,2}$,
            N. P. Vizarim$^{3}$, 
            C. J. O. Reichhardt$^2$, 
            P. A. Venegas$^4$,
            and C. Reichhardt$^2$}
            
\affiliation{$^1$ POSMAT - Programa de P\'os-Gradua\c{c}\~ao em Ci\^encia e Tecnologia de Materiais, S\~ao Paulo State University (UNESP), School of Sciences, Bauru 17033-360, SP, Brazil}

\affiliation{$^2$ Theoretical Division and Center for Nonlinear Studies, Los Alamos National Laboratory, Los Alamos, New Mexico 87545, USA}

\affiliation{$^3$ Department of Electronics and Telecommunications Engineering, S\~ao Paulo State University (UNESP), School of Engineering, S\~ao J\~oao da Boa Vista 13876-750, SP, Brazil}

\affiliation{$^4$ Department of Physics, S\~ao Paulo State University (UNESP), School of Sciences, Bauru 17033-360, SP, Brazil}

\date{\today}

\begin{abstract}
  Using atomistic simulations, we show that new types of skyrmion states called skyrmion molecular crystals and skyrmion superlattices can be realized on triangular substrates when there are two or three skyrmions per substrate minimum. We find that as a function of the magnetic field and substrate periodicity, a remarkably wide variety of ordered phases appear similar to those found in colloidal or Wigner molecular crystals, including ferromagnetic and herringbone states. The ability of the skyrmions to annihilate, deform, and change size gives rise to a variety of superlattice states in which a mixture of different skyrmion sizes and shapes produces bipartate or more complex lattice structures. Our results are relevant for skyrmions on structured triangular substrates, in magnetic arrays, or skyrmions in moir{\' e} materials.
\end{abstract}

\maketitle

{\it Introduction} 
Skyrmions are particle-like topologically protected
magnetic textures \cite{nagaosa_topological_2013,je_direct_2020}
that form a triangular lattice in the absence of a substrate
\cite{pfleiderer_skyrmion_2010,yu_real-space_2010}.
The skyrmions can also interact with a two-dimensional
triangular substrate created via nanostructuring techniques 
\cite{Fernandes18,Saha19,Arjana20,juge_helium_2021,zhang21,reichhardt_statics_2022,kern22}
or generated by the periodic magnetic field modulation originating from
a superconducting vortex lattice \cite{Petrovic21}.
Skyrmion states are also proposed to occur
in van der Waals or moir{\' e} materials,
where the skyrmions would also interact with
an underlying triangular substrate
\cite{hejazi_heterobilayer_2021,Akram21,sun23}.
Skyrmions might be expected to adopt the
symmetry of a triangular substrate since
it has the same symmetry as the substrate-free skyrmion lattice;
however, previous works on particle-like assemblies such as
superconducting vortices \cite{berdiyorov_vortex_2006,neal_competing_2007},
colloids \cite{reichhardt_novel_2002,brunner_phase_2002,agra_theory_2004,frey_melting_2005,thomas_spin_2007},
or Wigner crystals \cite{Reddy23,Li24}
interacting with triangular and square substrates
have shown that a wide variety of different ordered structures
can be stabilized when there are multiple particles per substrate minimum.
For colloidal
systems, such states are known as colloidal molecular crystals,
where $n$ colloids occupying each substrate minimum form an $n$-mer state such
as dimers for $n=2$ or trimers for $n=3$.
These $n$-mers exhibit orientational
orderings similar to those found in spin systems,
including antiferromagnetic, ferromagnetic, and herringbone states
\cite{reichhardt_novel_2002,brunner_phase_2002,agra_theory_2004,frey_melting_2005,thomas_spin_2007}.
Recently, Wigner molecular crystal states were proposed
\cite{Reddy23} and observed \cite{Li24}
for charge ordering in moir{\' e} systems where
each minimum captures multiple charges and there is orientational
ordering in adjacent minima.
For skyrmions on a triangular substrate, 
a much larger variety of ordered states should be possible
since the skyrmions can change
their size or shape and might also annihilate.

In this work we use atomistic simulations to investigate skyrmions interacting
with triangular substrates generated by periodic modulation of the material
anisotropy.
We focus on fillings of $n=2$ or 3 skyrmions per substrate minimum
as we vary the magnetic field and substrate spacing,
and show that a new class of skyrmion crystals
called skyrmion molecular crystals can be realized.
For $n=2$, skyrmion dimers appear and can order into
herringbone, ferromagnetic, and tilted states similar to those found for
colloidal molecular crystals
on triangular substrates.
We also find superlattice states composed of skyrmions of different sizes,
along with states
in which a portion of the skyrmions annihilate to create an $n=1$
commensurate filling.
For $n=3$, we observe aligned, tilted, and ferromagnetic
molecular crystals along with superlattice
states in which some of the skyrmions annihilate
and stripe ordering emerges.
For both $n=2$ and $n=3$,
we also find several states with elongated skyrmions or a combination of
circular and elongated skyrmions.

\begin{figure}[htbp]
    \centering
    \includegraphics[width=\columnwidth]{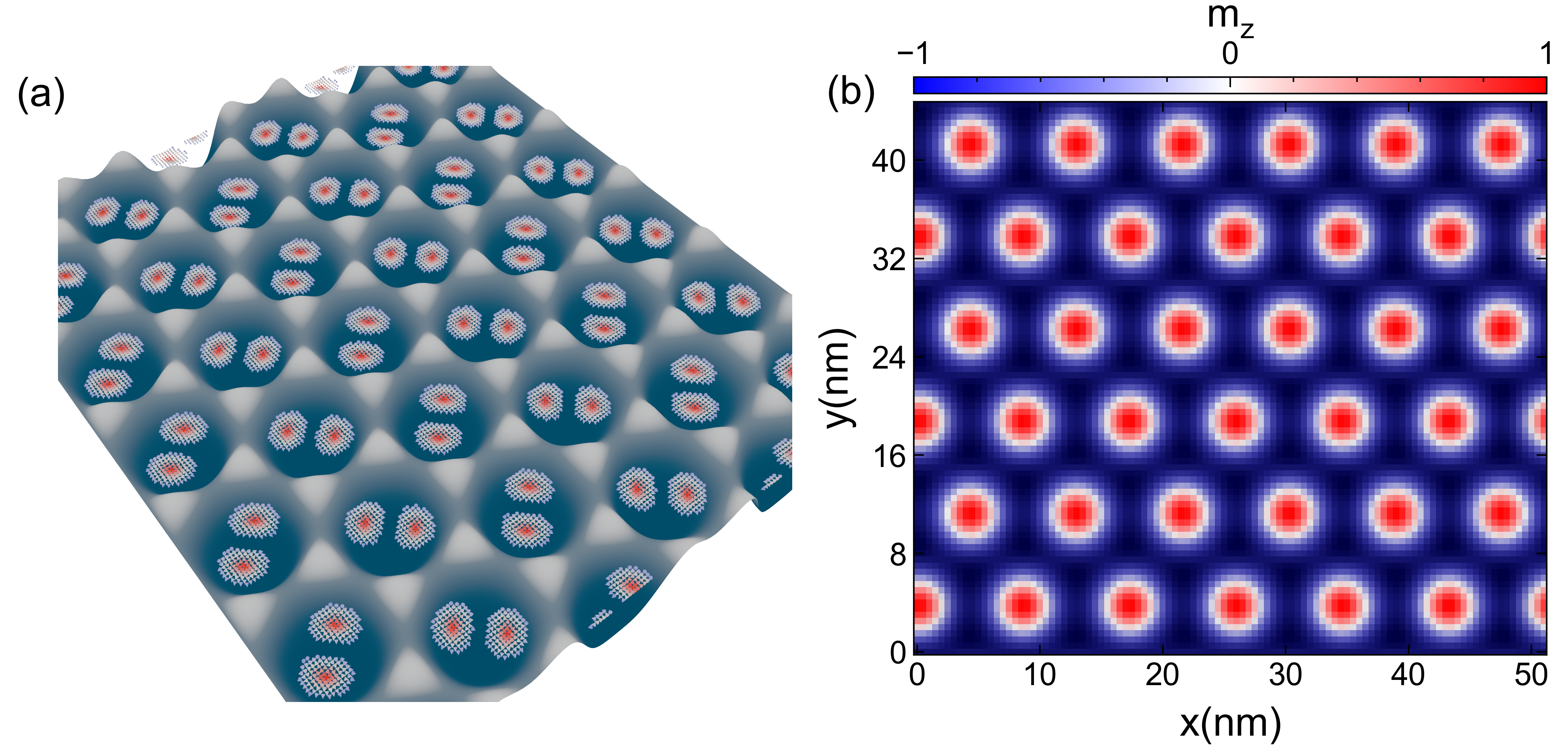}
    \includegraphics[width=\columnwidth]{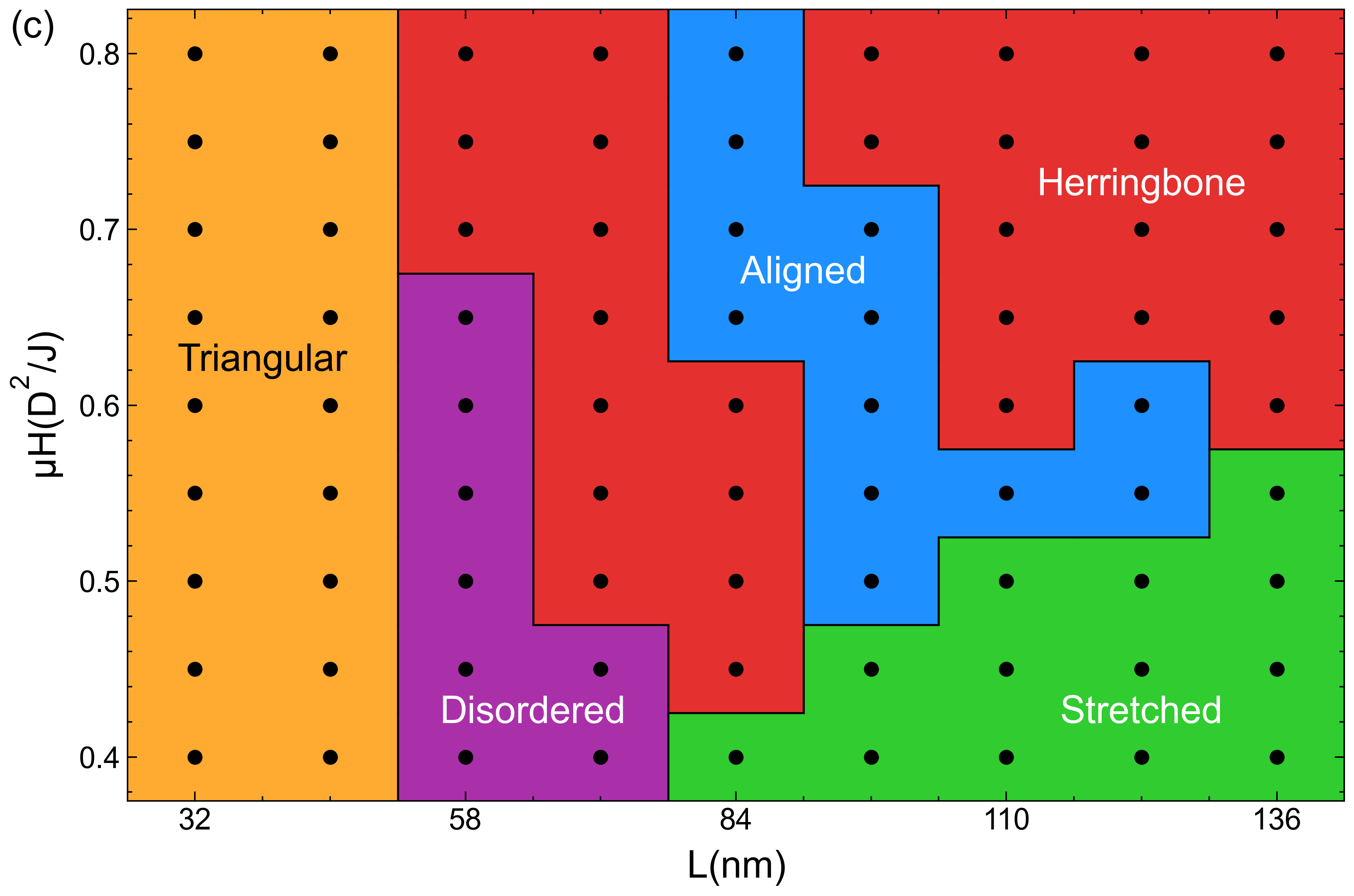}
\caption{
(a) Schematic of the system showing substrate minima (blue) and maxima
(white). Colored dots indicate the magnetic moments inside the skyrmions
for an $n=2$ filling.  
(b) Heightmap of the spatial variation of the magnetic moment $m_z$
for a system with $L=45$ nm and $\mu H = 0.4 D^2/J$.
The PMA is plotted as a transparency overlay, so that in nearly black
regions, the PMA is large and $m_z=-1$.
Half of the skyrmions annihilate, leaving behind
a commensurate one-to-one triangular skyrmion lattice.
(c) Phase diagram as a function of $\mu H$ vs $L$ for
the $n=2$ system, showing triangular [orange, Fig.~\ref{fig:1}(b)],
herringbone [red, Fig.~\ref{fig:2}(a,b)],
ferromagnetic or aligned [blue, Fig.~\ref{fig:2}(c,d)],
stretched [green, Fig.~\ref{fig:2}(e,f)], and
disordered [purple, Fig.~\ref{fig:2}(g,h)] phases.
}
\label{fig:1}
\end{figure}

{\it Simulation---} 
We model Néel skyrmions
in a ultrathin ferromagnetic
sample using an atomistic
model \cite{evans_atomistic_2018} that describes the dynamics of
individual atomic magnetic moments.
The system size is $(2/\sqrt 3)L\times L$
with periodic boundary conditions along
the $x$ and $y$ directions.
The triangular substrate, shown schematically in
Fig.~\ref{fig:1}(a) at the $n=2$ filling, is created by modulating the
perpendicular magnetic anisotropy (PMA):

\begin{equation}\label{eq1}
    K(x, y)= \sum_{i=1}^{3}\frac{K_0}{2}\left[\cos\left(\frac{2\pi n}{L}b_i\right) + 1\right]
\end{equation}
where $K_0$ is the depth of the PMA modulation,
$b_i=x\cos(\theta_i)-y\sin(\theta_i)+L/2N_m$,
$N_m=6$, $\theta_1=\pi/6$, $\theta_2=\pi/2$, and $\theta_3=5\pi/6$.
The magnetic field is applied perpendicular to the sample
surface along the negative $z$ direction.

The Hamiltonian governing the
$T=0$ atomistic dynamics of an underlying square arrangement of magnetic moments
with lattice constant $a=0.5$ nm
is given by
\cite{evans_atomistic_2018, iwasaki_universal_2013, iwasaki_current-induced_2013}:

\begin{align}\label{eq2}
  \mathcal{H}=&-\sum_{i, j\in n.n.}J_{ij}\mathbf{m}_i\cdot\mathbf{m}_j
                -\sum_{i, j\in n.n.}\mathbf{D}_{ij}\cdot\left(\mathbf{m}_i\times\mathbf{m}_j\right)\\\nonumber
                &-\sum_i\mu\mathbf{H}\cdot\mathbf{m}_i
                -\sum_{i} K(x_i, y_i)\left(\mathbf{m}_i\cdot\hat{\mathbf{z}}\right)^2\\\nonumber
\end{align}
The first term on the right side is the exchange interaction with
coefficient $J_{ij}=J$ between nearest neighbors.
The second term is the interfacial Dzyaloshinskii–Moriya
interaction, where
$\mathbf{D}_{ij}=D\mathbf{\hat{z}}\times\mathbf{\hat{r}}_{ij}$
and $\mathbf{\hat{r}}_{ij}$ is the
unit distance vector between sites $i$ and $j$.
The third term is the Zeeman interaction with an applied external magnetic
field $\mathbf{H}$.
Here $\mu=\hbar\gamma$ is the magnitude of the magnetic moment
and $\gamma=1.76\times10^{11}$T$^{-1}$s$^{-1}$ is the electron
gyromagnetic ratio. The last term represents the 
perpendicular magnetic anisotropy (PMA) of the sample, where
$\mathbf{r}_i=x_i\mathbf{\hat{x}}+y_i\mathbf{\hat{y}}$.
Since we are considering ultrathin films, long-range dipolar interactions
are small enough that they can be neglected \cite{paul_role_2020}.

The time evolution of the magnetic moments is obtained from
the LLG
equation \cite{seki_skyrmions_2016,gilbert_phenomenological_2004}:

\begin{equation}\label{eq3}
    \frac{\partial\mathbf{m}_i}{\partial t}=-\gamma\mathbf{m}_i\times\mathbf{H}^\text{eff}_i
                             +\alpha\mathbf{m}_i\times\frac{\partial\mathbf{m}_i}{\partial t} \ ,
\end{equation}
where $\gamma$ is the electron gyromagnetic ratio,
$\mathbf{H}^\text{eff}_i=-\frac{1}{\hbar\gamma}\frac{\partial \mathcal{H}}{\partial \mathbf{m}_i}$
is the effective magnetic field including all interactions from
the Hamiltonian, and $\alpha$ is the Gilbert damping.
Equation~\ref{eq2} is integrated using
a fourth order Runge-Kutta method, and each simulation spans a time
of 20 ns.
We fix
$\alpha=0.3$, $K_0=0.1J$, and $n=6$. The material parameters are $J=1$ meV and
$D=0.5J$.
For the initial conditions, the sample is seeded with a number of skyrmions
equal to either an $n=2$ or $n=3$ filling.
We image $m_z$ in real space and also compute
the perpendicular spin structure factor \cite{mazurenko_role_2016}:
\begin{equation}\label{eq4}
    S_\perp\left(\mathbf{q}\right)=\left|\sum_i m_i^x \text{e}^{-i\mathbf{r}_i\cdot \mathbf{q}}\right|^2+\left|\sum_i m_i^y \text{e}^{-i\mathbf{r}_i\cdot\mathbf{q}}\right|^2
\end{equation}
where
$\mathbf{q}$
is the wave-vector. Since the spin structure factor has
arbitrary units, we plot $\widetilde{S}_\perp$, the
normalized value of $S_\perp$ over the range $[0, 1]$.

\begin{figure}[htbp]
    \centering
    \includegraphics[width=\columnwidth]{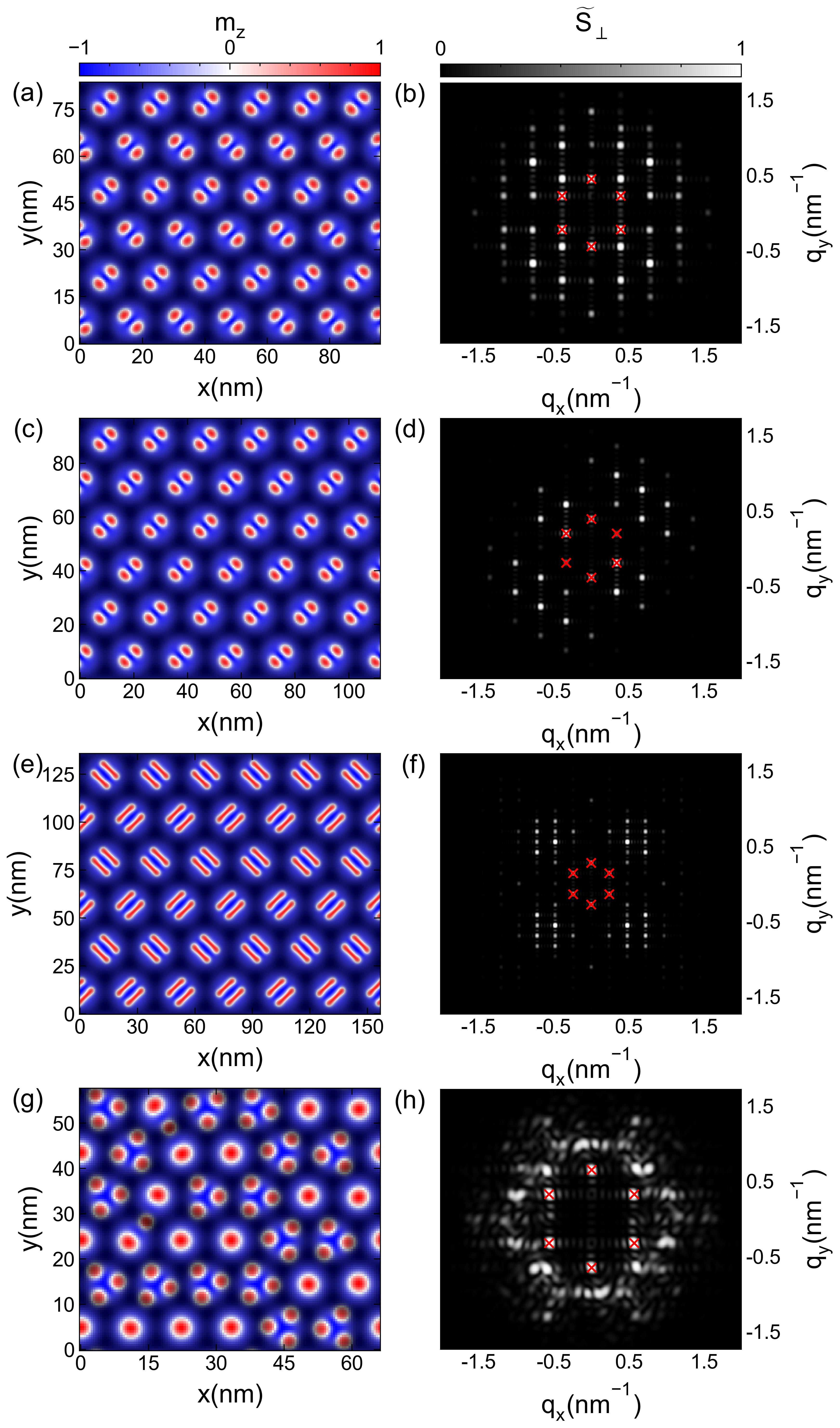}
\caption{(a,c,e,g) Real space heightmap images of $m_z$
for the $n=2$ phases from Fig.~\ref{fig:1}(c)
colored as in Fig.~\ref{fig:1}(b).
(b,d,f,h) The corresponding
normalized perpendicular spin structure $\widetilde{S}_{\perp}$, where
the red $\times$ symbols
indicate the positions
of the trivial peaks produced by the substrate lattice spacing.
The positions of these peaks changes when the value of $L$ changes.
(a,b) At $\mu H=0.55D^2/J$ and $L=84$ nm,
the skyrmions form a dimer lattice with herringbone order.
(c,d) At $\mu H=0.55D^2/J$ and $L=84$ nm,
an aligned dimer or ferromagnetic state appears.
(e,f) In the stretched phase at
$\mu H=0.45D^2/J$ and $L=136$ nm, the skyrmions elongate in order to span
the potential minima.
(g,h) A disordered state
at  $\mu H=0.45D^2/J$
and $L=58$ nm where there is
a mixture of monomers and trimers.
	} 
	\label{fig:2}
\end{figure}

{\it $n=2$ Dimers---}
We first consider samples initialized with $n=2$ and
conduct a series of simulations for varied
magnetic field $\mu H$ and sample size $L$.
In Fig.~\ref{fig:1}(c), we plot a phase diagram as a function of
$\mu H$ versus $L$ for the $n=2$ system.
For $L < 58$nm, the minima are too small to accommodate two skyrmions
and half of the skyrmions annihilate in order to form a one-to-one
commensurate triangular state, as illustrated in Fig.~\ref{fig:1}(b)
for $L=45$ nm and $\mu H=0.4 D^2/J$.
In this state, the only peaks that appear in the
perpendicular spin structure factor
$\widetilde{S}_{\perp}$ (not shown) are those falling at the trivial
wavevectors corresponding to the substrate lattice spacing.
In the herringbone state, which appears in two separated regions of the
phase diagram, the dimers in each minimum form tilted rows as
shown in
Fig.~\ref{fig:2}(a)
at $L=84$ nm and $\mu H=0.55D^2/J$.
This structure is similar to that found for colloidal
molecular crystals on a triangular substrate
\cite{reichhardt_novel_2002,frey_melting_2005,thomas_spin_2007};
however, due to their internal degrees of freedom, the skyrmions
are slightly distorted in the herringbone state.
The corresponding $\widetilde{S}_{\perp}$
is plotted
in Fig.~\ref{fig:2}(b). In addition to the trivial peaks, marked with
red $x$'s, we observe peaks at larger wavevectors corresponding to
the substructure inside each potential minimum.
In between the two herringbone regimes in Fig.~\ref{fig:1}(c), we find
an aligned or ferromagnetic phase
corresponding to
a state in which all of the dimers spontaneously break spatial symmetry
and align in a single direction, as illustrated in 
Fig.~\ref{fig:2}(c) for $\mu H=0.55D^2/J$
and $L=97$ nm. The spontaneous symmetry breaking produces
asymmetric peaks in $\widetilde{S}_{\perp}$, as shown in Fig.~\ref{fig:2}(d).
The stretched state in Fig.~\ref{fig:1}(c) is a variant on the
herringbone state and is illustrated in
Fig.~\ref{fig:2}(e)
at $\mu H=0.45D^2/J$ and  $L=136$ nm.
Each skyrmion elongates in order to span the full width of the substrate
minimum, and this can be viewed as the splitting of each skyrmion into
two merons, making the state equivalent to a filling of four merons
per substrate minimum. Peaks in $\widetilde{S}_\perp$ extend out to
larger wavevectors in Fig.~\ref{fig:2}(f) due to the small spacing between
the merons.
In addition to the ordered $n=2$ states described above, Fig.~\ref{fig:1}(c)
contains an extended disordered regime. As illustrated in
Fig.~\ref{fig:2}(g) at $\mu H=0.45D^2/J$ and $L=58$ nm, this state is
composed of a mixture of monomers and trimers, and the size of the skyrmions
ranges from very large in the monomer state to very small in the trimer
state. On average, there are still two skyrmions per substrate minimum.
The corresponding $\widetilde{S}_\perp$
in Fig.~\ref{fig:2}(h) has peaks only at the trivial positions, indicating
a lack of long-range ordering of the $n$-mers.
Although the herringbone and aligned states are found in colloid systems
and Wigner molecular crystals, the stretched phase and disordered phases
can only occur due
to the ability of the skyrmions to change size and shape.

\begin{figure}[htbp]
    \centering
    \includegraphics[width=\columnwidth]{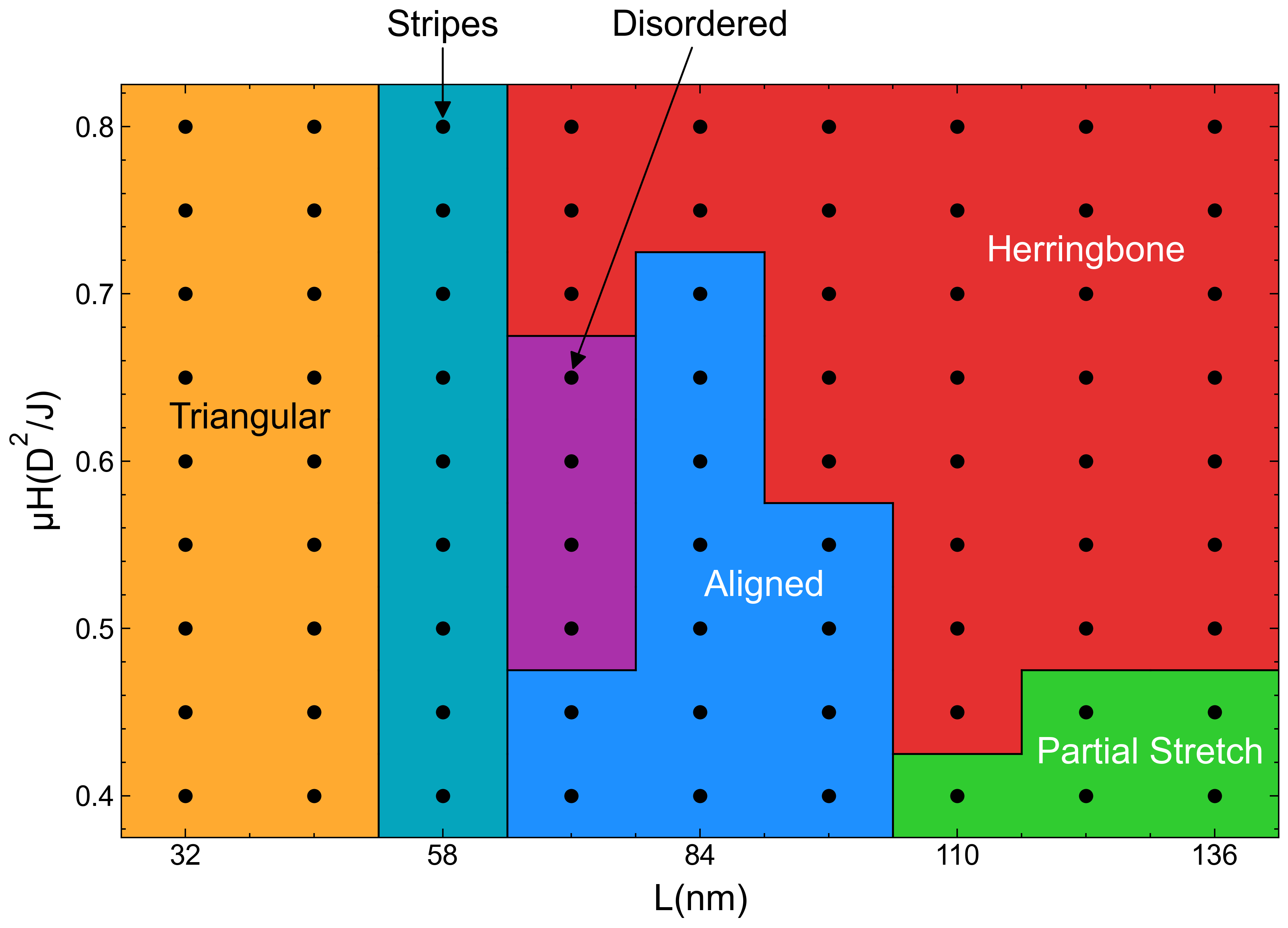}
    \caption{
Phase diagram as a function of $\mu H$ vs $L$ for
the $n=3$ system, showing      
triangular [orange, Fig.~\ref{fig:1}(b)],
stripe superlattice [teal, Fig.~\ref{fig:4}(a,b)],
aligned [blue, Fig.~\ref{fig:4}(c,d)],
herringbone [red, Fig.~\ref{fig:4}(e,f)],
and partially stretched [green, Fig.~\ref{fig:4}(g,h)] phases.
In the purple region
there is a disordered phase consisting of
an aligned phase in which
some of the trimers have been flipped.
}
\label{fig:3}
\end{figure}

\begin{figure}[htbp]
    \centering
    \includegraphics[width=\columnwidth]{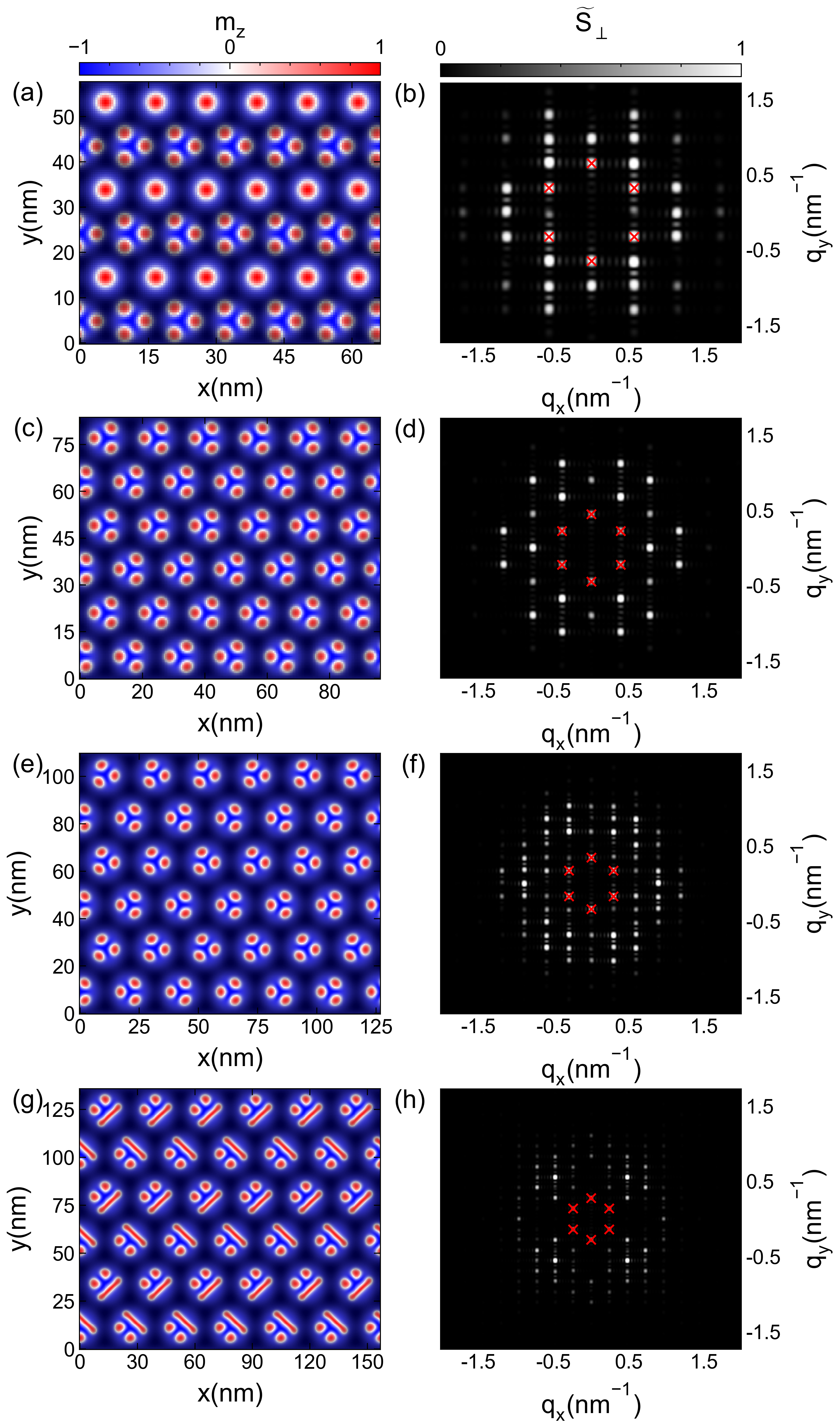}
\caption{
(a,c,e,g) Real space heightfield images of $m_z$ colored as in      
Fig.~\ref{fig:1}(b) and (b,d,f,h) $\widetilde{S}_\perp$ for the
$n=3$ phases from Fig.~\ref{fig:3}. The red $\times$ symbols in
$\widetilde{S}_\perp$ indicate the positions of the trivial peaks
produced by the substrate lattice spacing.
(a,b) The superlattice stripe phase of monomers and trimers
at $\mu H=0.5D^2/J$ and $L=58$ nm.
(c,d) The aligned trimer or ferromagnetic state
at $\mu H=0.5D^2/J$ and $L=84$ nm.
(e,f) A herringbone trimer state
at $\mu H=0.5D^2/J$
and $L=110$ nm.
(g,h) A partially stretched phase
at $\mu H=0.4D^2/J$ and $L=136$ nm, where one of the skyrmions splits into
two merons and the other two skyrmions form a dimer structure.
}
\label{fig:4} 
\end{figure}

{\it $n=3$ Trimers--- }
In Fig.~\ref{fig:3} we construct a phase diagram as a function of
$\mu H$ versus $L$ for a system initialized with $n=3$.
When $L < 58$ nm, two-thirds of the
skyrmions annihilate
and the system forms a triangular monomer lattice with one-to-one matching
to the substrate, similar to the state illustrated in
Fig.~\ref{fig:1}(b).
At $L = 58$ nm we find 
an ordered superlattice stripe
state, shown in Fig.~\ref{fig:4}(a)
at $\mu H=0.5D^2/J$ and $L=58$ nm, where one-third of the skyrmions have
annihilated.
Every other row contains large skyrmion monomer states, while the remaining rows
contain aligned trimers of smaller skyrmions.
The corresponding $\widetilde{S}_\perp$ in Fig.~\ref{fig:4}(b)
has peaks extending along
$q_y$, which reflects the stripe ordering.
In the aligned or ferromagnetic state, shown in
Fig.~\ref{fig:4}(c) 
at $\mu H=0.5D^2/J$ and $L=84$ nm,
all of the trimers have the same orientation. This state has been observed
for colloids on triangular substrates in simulation \cite{reichhardt_novel_2002}
and experiment \cite{brunner_phase_2002}
as well as for Wigner crystal
molecules \cite{Li24}.
Secondary triangular peaks appear in the corresponding $\widetilde{S}_\perp$
in Fig.~\ref{fig:4}(d).
In the herringbone state, illustrated
in Fig.~\ref{fig:4}(e,f)
at $\mu H=0.5D^2/J$ and $L=110$ nm,
the orientation of the trimers alternates from one row to the next.
We find an unusual partially stretched phase in which one skyrmion splits
into two merons while the other two skyrmions form a dimer, as shown
in Fig.~\ref{fig:4}(g,h) at $\mu H=0.4D^2/J$ and $L=136$ nm.
The resulting structure has a facelike appearance, and the faces show
an alternating orientational ordering.
Finally, we observe a disordered phase that consists of an aligned trimer
state of the type shown in Fig.~\ref{fig:4}(c) but with the orientation
of some of the trimers flipped.

We expect that 
additional phases will occur
for other fillings, fields, lattice sizes, and different lattice
geometries such as square, rectangular, or quasiperiodic.
We only considered integer fillings,
but other superlattice orderings could appear for
filings of $n=3/2$, $n=5/2$, $n=4/3$, and higher order fillings.
Our results suggest that highly symmetrical states will
be the most stable.
Application of external driving could produce interesting effects,
such as by producing a biasing field that aligns the $n$-mers with the
driving direction, as observed
for colloidal molecular crystals
\cite{Reichhardt09,Shawish11}.
We also expect that interesting resonance modes would occur
for the different states under ac excitation
or time dependent magnetic fields.
For fillings just outside of commensuration,
there could be kink or anti-kink flow
in the dimer or trimer lattices.
Colloidal molecular crystals
are known to have multiple melting transitions
\cite{reichhardt_novel_2002,brunner_phase_2002,
agra_theory_2004,frey_melting_2005},
and the skyrmion molecular crystals should show similar melting
transitions as well as possible additional transitions caused by
thermal deformations of the skyrmions, making it possible to
observe orientationally disordered states that do not occur in the
particle-based molecular crystal systems.
The trimer and dimer states we observe could be useful for creating
new types of memory devices
where the orientation of the $n$-mer can be used to store a bit of information.
Beyond skyrmions, our results are relevant for other types
of magnetic or charge bubble-like textures interacting with a triangular
substrate,
which could include liquid
crystal skyrmions, magnetic vortices,
charge ordered states in quantum Hall systems, and certain emulsions.

{\it Summary---}
Using atomistic dynamics,
we have shown that a new class of skyrmion structures 
called skyrmion molecular crystals and
skyrmion molecular superlattice
states can be realized for skyrmions
on a triangular substrate when there are
$n=2$ or $n=3$ three skyrmions per substrate minimum.
For $n=2$ 
we find ferromagnetic and herringbone
states, a disordered state where the skyrmions have multiple sizes,
and an ordered stretched or meron lattice phase.
For small lattice spacings, half of the skyrmions annihilate and
the skyrmions form a commensurate triangular lattice with a one-to-one
matching to the substrate.
For $n=3$ we find aligned and herringbone trimer
states, stripe superlattices containing a mixture of monomers and trimers,
and a partially stretched state where
one skyrmion stretches into two merons
and the other skyrmions remain
circular.
Several of the skyrmion molecular crystal
states we observe are similar to the colloidal
and Wigner molecular crystals found for
triangular substrates;
however, the stretched, partially stretched, disordered and superlattice states
can only occur due to the ability of the skyrmions to change their
size and shape, and
are absent
in particle-based models.
Our results open the possibility of a new class of
skyrmion states
that can be realized in nanostructured skyrmion systems or for skyrmions
in moir{\' e} systems.
The orientation of individual $n$-mers could be used as a memory
storage mechanism.
Our results should
also be relevant to other types 
of particle like textures that can change size
and/or deform when coupled to a triangular lattice.

\acknowledgments
This work was supported by the US Department of Energy through the Los Alamos National Laboratory. Los
Alamos National Laboratory is operated by Triad National Security, LLC, for the National Nuclear Security
Administration of the U. S. Department of Energy (Contract No. 892333218NCA000001). 
J.C.B.S acknowledges funding from Fundação de Amparo à Pesquisa do Estado de São Paulo - FAPESP (Grant 2023/17545-1).
We would like to thank Dr. Felipe F. Fanchini for providing the computational resources used in this work. 
These resources were funded by the Fundação de Amparo à Pesquisa do Estado de São Paulo - FAPESP (Grant: 2021/04655-8).

\bibliography{mybib}

\end{document}